\documentclass[12pt,preprint]{aastex}
\usepackage{natbib}
\def\kms{\, \rm{km}\,\rm{s}^{-1}}

\shorttitle{Racusin et al.}
\shortauthors{Judith Racusin}

\begin{document}

\title{X-ray Evolution of SNR 1987A: The Radial Expansion}
\author{Judith L. Racusin\altaffilmark{1}, Sangwook Park\altaffilmark{1},
  Svetozar Zhekov\altaffilmark{2,3},
  David N. Burrows\altaffilmark{1}, Gordon P. Garmire\altaffilmark{1},
  Richard McCray\altaffilmark{2}}
\altaffiltext{1}{Department of Astronomy \& Astrophysics, The Pennsylvania
  State University, University Park, PA 16802}
\altaffiltext{2}{JILA, University of Colorado, Box 440, Boulder, CO, 80309}
\altaffiltext{3}{On leave from Space Research Institute, Sofia, Bulgaria}

\begin{abstract}
  We present the evolution of the radial expansion of SNR 1987A as
  measured using {\it Chandra} X-ray observations taken over the last 10 
  years.  To characterize the complex structure of the remnant and isolate
  the expansion measurement, we fit the
  images to several empirical models including: a simple circular
  torus, a torus with bilateral lobes, and a 
  torus with four tangentially extended lobes.  We discuss the results of
  this measure in the context of the overall evolution of the supernova
  remnant, for which we believe we have measured the end of the free
  expansion phase and its transition to the adiabatic phase (at least
  along the equatorial ring).  The timing of
  this event is in agreement with early predictions of the remnant evolution.
  
\end{abstract}

\section{Introduction\label{sec:intro}}
Supernova 1987A is the nearest and consequently the most extensively
studied supernova (SN) in modern 
astronomy.  While it has led to important insights into progenitor systems
and supernova evolution, it is by no means a typical characteristic event.
The progenitor was seen in archival plates as blue supergiant {\it Sk
  -69 202} \citep{white87} rather than the expected red supergiant
progenitor of a Type II SN.  The extent of the stellar winds from 
both the red and blue supergiant phases suggest that the system transitioned
from a red supergiant sometime in the last $\sim 40,000$ years
\citep{arnett89}.  This transition may explain some of the unusual
properties of the supernova remnant (SNR). 

The ring-like structure visible and evolving over time in X-ray, optical,
and radio observations was first discovered to have ``clumpy'' structure
using the high-resolution Hubble Space Telescope images \citep{plait95}.
The many spots seen in the optical appear to correspond to the less resolved
soft X-ray spots.  The X-ray and radio image structures are much simpler than the
complex triple-ring structure seen in optical observations
\citep{burrows95}; the X-rays and radio emission are only detected 
from the central inner ring.  This ring appears to be a structure of hot
gas that is flattened into a round disk.  It appears elliptical due to
its inclination, is gradually brightening in all bands, shows increasing
structural details, and is expanding radially.  The optical emitting region of
the inner ring is slightly outside that of the X-ray and radio emitting
regions.   
The soft X-ray emission from SNR 1987A is thought to be produced by shocked
hot gas that lies between the forward and reverse shocks of the SN blast
wave, with the localized spots of emission around the inner ring being due
to regions of enhanced density protruding from the inner surface of the
circumstellar ring \citep{park02,sugerman02}.  The inner and outer rings are
thought to be produced by asymmetric stellar winds shed at different stages
in the evolution of the massive progenitor star \citep{luo91,cheva95}.  

\begin{figure}
    \begin{center}
      \includegraphics[scale=0.8]{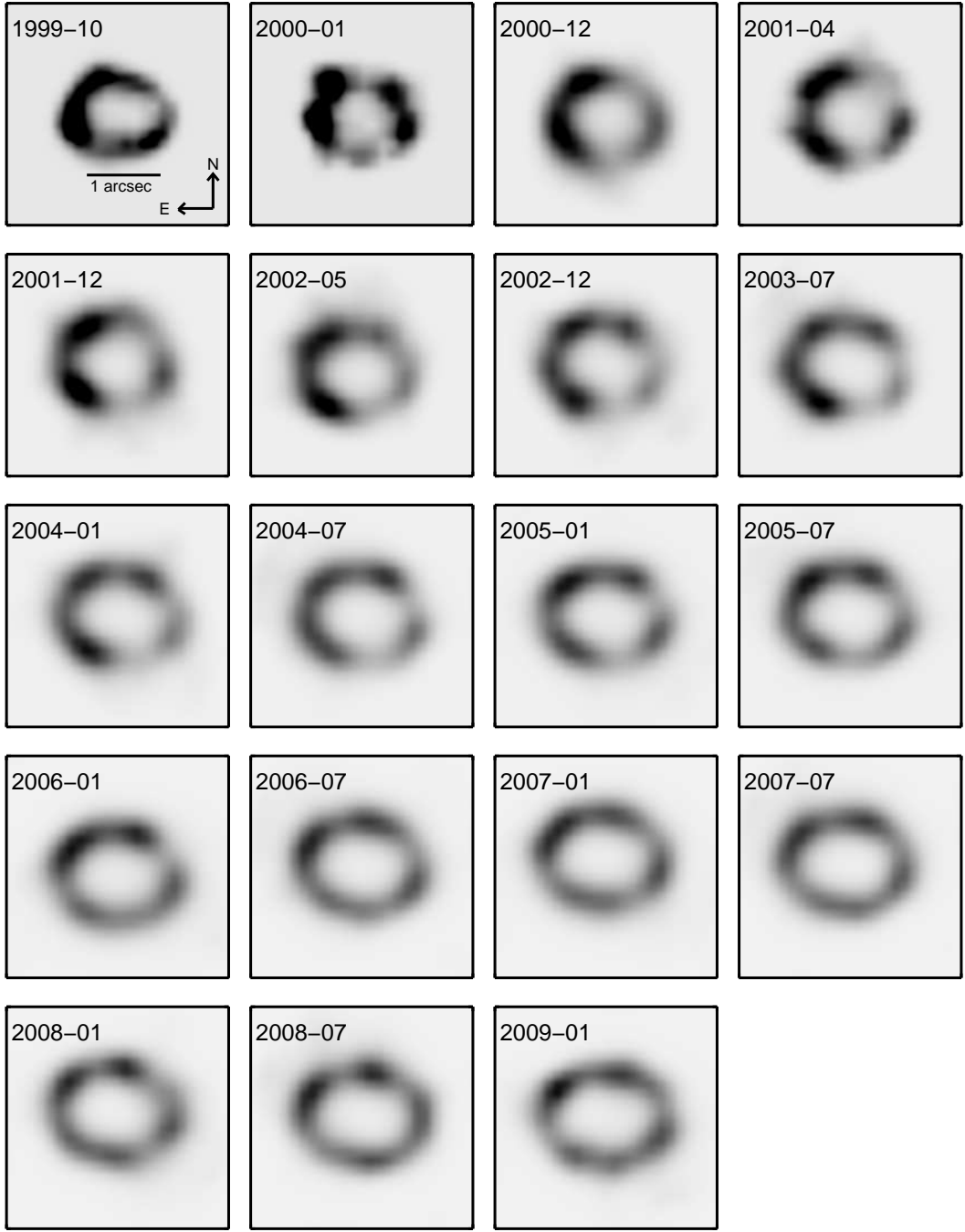}
      \figcaption{Broadband ($0.3-8.0$ keV) images of SNR 1987A from each
        {\it Chandra} observation taken on date (YYYY-MM) as labeled.
        The images are all the same spatial scale with slight uncertainty in
        rotations and centers due to unknown absolute astrometry on these
        scales.  The images are scaled arbitrarily to show the
        morphological details of the evolving remnant.  The flux has
        actually increased dramatically
        over the course of this monitoring campaign, as shown in Figure
        \ref{fig:xlc}. \label{fig:87a_images}}
    \end{center}
  \end{figure}

The soft X-ray emission has been increasing in
intensity as a result of the blast wave shocking and continually sweeping up
the surrounding circumstellar material.  The  
structure of the remnant has evolved from emission largely contained in these
small regions to overall brightening of the ring during 
the progressive observations of the resolved remnant.  As the blast wave
lights up the dense inner ring, the X-ray spectrum may become dominated by
the density structure of the ring itself \citep{park06a}.  Figure
\ref{fig:87a_images} shows the $0.3-8.0$\ keV {\it Chandra} images at each
epoch (Table \ref{tab:obs_detail}), showing how the X-ray emission has transitioned
in recent years from being 
dominated by the shocked gas in the dense hot spots to being dominated by the
gas between the forward and reverse shocks \citep{mccray03}.

\begin{figure}
    \begin{center}
      \includegraphics[scale=0.7,angle=90]{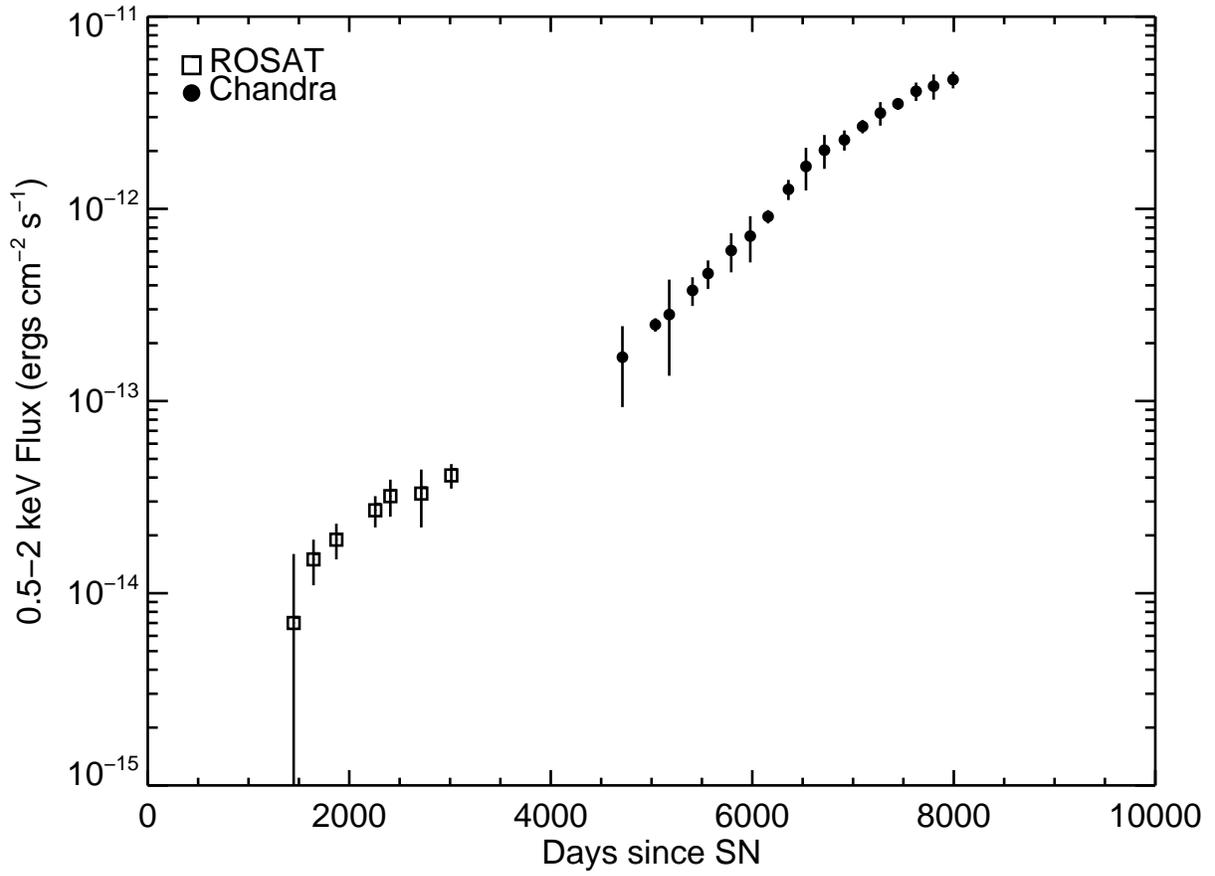}
      \figcaption{Soft X-ray ($0.5-2$ kev) light curve from ROSAT and {\it
          Chandra} observations.
        \label{fig:xlc}}
    \end{center}
  \end{figure}

The {\it ROSAT} X-ray light curve increased linearly for approximately
10 years after the SN explosion.  Later {\it Chandra} observations
indicated that there was a deviation beginning in $\sim1997$ ($\sim 3700$
days after the SN) 
from the {\it ROSAT} linear extrapolation of the flux, which results
in a nearly exponential increase in the soft X-ray light curve (Figure
\ref{fig:xlc}).  This transition was temporally
coincident with the appearance of the first optical spot
\citep{park02,sonneborn98}.  This rapid flux increase was followed by an 
upturn in the X-ray light curve in $2004$ ($\sim6200$ days after the
SN), suggesting that the blast wave had reached the main body of the
inner ring. This brightening was accompanied by substantial softening of
the overall X-ray spectrum, brightening of hot spots in the ring,
filling in of the X-ray ring, and a decrease in radial expansion velocity of
the ring. 
  
In this paper we describe a new method for measuring the size of the remnant
from the X-ray images, and use that to put
it in the context of the structure and overall evolution of the system.  We 
describe in \S\ref{sec:predict} the predictions in the literature prior to
our {\it Chandra} campaign, the observations in \S\ref{sec:obs}, our radial
expansion measurement and image models in \S\ref{sec:radexp}, the results in
\S\ref{sec:results}, the discussion in 
\S\ref{sec:sims}, and our conclusions in \S\ref{sec:conclusions}.
  
\section{Predictions \label{sec:predict}}
Early observations of the circumstellar ring and measurements of the ejecta
velocities prompted predictions for when the blast wave would encounter the
inner ring \citep{cheva89,luo91}.  This encounter was predicted to occur
between 8 and 16 years after the SN and would result in dramatic
re-brightening as a result of the strong shocks produced by the blast wave
interacting with the inner ring.  This
prediction was later revised when \cite{cheva95} suggested that an H II
region may exist inside the inner circumstellar ring.  It is needed to
reconcile the expected low mass loss rate ($\dot{M}\sim7.5\times 10^{-8}\,
\rm{M_{\odot}}\, \rm{yr}^{-1}$) of the blue supergiant progenitor with the
early radio free-free absorption and the X-ray emission levels.  The H II
region was probably formed in material swept up by the red supergiant wind
that was ionized by the blue supergiant star.  This has the effect of
delaying the date of the encounter with the inner ring to year $2005\pm3$ 
because the blast wave is slowed as it swept up this additional material.

\cite{bork} modeled the SN blast wave impact with the inner circumstellar
ring and the resulting reverse shock.  They predicted that the impact would
be characterized by a gradual increase in X-ray luminosity.  As the shock
front engulfed the inner ring, it was expected to gradually brighten by as
much as 3 orders of magnitude in the optical, UV, and X-ray.  The X-ray
spectrum would also change depending on the detailed shock structure of the
interaction with the inner ring.  The overall soft X-ray flux evolution is
consistent with these predictions, showing a steep increase in the light
curve since $\sim 2000$ (see Figure \ref{fig:xlc}). However, these
predictions were made  
before the emergence of the first optical spot, and therefore could not have
accounted for how the higher density protrusions would affect the
observational behavior.  

After $\sim1997$ ($\sim 3700$ days after the SN) the soft X-ray light
curve began to deviate from the linear increase observed for the prior 10
years.  This brightening was initially confined to the localized hot spots
evident in HST images in the northeastern and southeastern regions of the
inner ring.  As more optical hot spots appeared, the X-ray emission also
spread around the ring and the X-ray luminosity increased dramatically.
The brightening of the entire ring of SNR 1987A appears to have begun in
early $2004$ and the events predicted are underway. This is evident in the
brightening in X-rays \citep{park05}, infrared \citep{bouchet06}, and
optical \citep{mccray05}, the continuous softening of the X-ray
spectrum \citep{park06b,zhekov05}, deceleration of the radial expansion of the
SNR \citep{park04,zhekov05}, and slower shock velocities from line broadening
\citep{michael02,dewey08,zhekov09}.

\section{{\it Chandra} Observations and Data Reduction \label{sec:obs}}
We have observed SNR 1987A with {\it Chandra} on 23 occasions between
1999 October and 2009 January, and our monitoring observations continue at
approximately 6 month intervals. Eighteen of the observations were carried out
using the Advanced CCD Imaging Spectrometer (ACIS) back-illuminated S3
CCD.  ACIS alone provides high-resolution images with moderate
non-dispersive energy resolution.  When used in conjunction with one of the
gratings, either the High Energy Transmission Grating (HETG) or the Low
Energy Transmission Grating (LETG), ACIS provides high quality spectra
and an undispersed zeroth-order image with moderate energy resolution.  Six
of our observations used the gratings: HETG (1999 October and 2007 April,
2008 July, 2009 January, 
\citep{burrows00,michael02,dewey08}) and LETG (2004 September and 2007
September, \citep{zhekov05,zhekov06,zhekov09}).  In the standard LETG configuration, the
long-wavelength oxygen lines are positioned on the sensitive ACIS-S3 chip,
which consequently places the zeroth order image 1.5\arcmin\ off-axis, blurring it
by {\it Chandra}'s off-axis Point Spread Function (PSF).  This blurring
degrades the image, and we therefore 
exclude the LETG observations from our measurement of the radial expansion
rate.  The HETG, however, does not suffer from this effect.  The differences
in the images between the bare-ACIS and the zeroth-order HETG have not been
calibrated, though no significant systematic effects (as seen with the LETG)
are expected.  The first HETG observation is also our first monitoring
observation, therefore it is an important reference point for our analysis.
The deep HETG observation taken in 2007 was not part of our regular
monitoring observations and was performed in between regularly scheduled
monitoring observations.  As the remnant brightens, we have recently taken
calibration observations comparing the bare-ACIS and HETG zeroth-order
images (in 2008-7 and 2009-1).  In order to avoid
photon pipe-up, we will transition future observations to HETG only.  These
closely spaced HETG observations to monitoring ACIS observations revealed 
consistent radii between the ACIS-only and HETG measurements.  They do not
add new information to this study and therefore do not have a strong affect
on the radial expansion velocity fits.  We exclude all gratings observations
except the 1999 October HETG observation from the rest of radial expansion
analysis presented in this paper. The details of the 19 {\it Chandra}
observations used in this work are listed in Table \ref{tab:obs_detail}. 

The technique used to create our images is the same as was used in our
previous works \citep{burrows00,park02,park04,park06a,park06b}.  The event
data used to create the images were filtered to use the energy range $0.3-8$
keV, and ASCA grades 02346.  Flaring pixels were removed.  We corrected for
the charge transfer inefficiency (CTI) using the methods developed by
\cite{townsley}.  The small angular extent of the remnant ($\sim1\farcs 6$)
makes any image analysis difficult even with the superb spatial resolution of
{\it Chandra} (ACIS CCD pixel size $\sim0\farcs 492$).  To better sample the
image, we applied the sub-pixel resolution method developed by
\cite{tsunemi} and \cite{mori}.  This technique uses the feature that
information can be obtained on the location within the
pixel where the photon was absorbed by 
examining the distribution of the events split between pixels.  The
morphology of the splitting can be traced to a centroid within the pixel for 
events spread across more than 2 pixels.  This technique improves the angular
resolution of the images by $\sim10\%$.

Several other factors also help to improve the image resolution with the use
of the sub-pixel resolution technique.  The choice of using one of the ACIS
back-illuminated CCDs leads to a larger fraction of split-events.  The back
illuminated chips are also less affected by the poor charge transfer
inefficiency for observations obtained after the particle damage in
September 1999 \citep{townsley}.  Another source of better resolution comes
from the intentional dithering motion of the observatory which causes a
target image to move across the CCD surface forming a Lissajous pattern.
This leads to more opportunities of forming split-pixel events in different
areas of the image. Standard processing adds a randomization of event
positions within each pixel to prevent marked pixelization of the images.
We turn off this randomization in the data processing to prevent any loss of
positional information.

To further improve the effective angular resolution of the images, we
deconvolve them with a maximum likelihood algorithm \citep{richardson, lucy}
using the on-axis detector PSF.  We used a scale of
$0\farcs 125$ sky pixels
and then smoothed the images by convolving with a Gaussian ($\sim0\farcs 1$
FWHM).  The resulting $0.3-8.0$ keV broadband images of SNR 1987A from each
of the epochs of our observations are shown in Figure \ref{fig:87a_images}.

\section{Radial Expansion \label{sec:radexp}}
Previous works to model the X-ray radial expansion rate of SNR 1987A
\citep{park02,park04} used a simple Gaussian model to fit the radially
averaged profile.  This approach yielded an expansion velocity estimate of
$4000-5000 \kms$.  The utility of this simple method was limited by the loss
of information on the detailed features such as the lobes, the overall
asymmetric intensity, as well as ellipticity and inclination of the inner
ring.  To more accurately quantify the radial expansion, we developed a more
realistic image model to estimate the SNR's
radius.

The goal for our empirical models is to characterize the {\it Chandra} X-ray
images in a more realistic way and in a similar spirit as those of the radio
studies 
\citep{gaensler,manch02,manch05,ng08} in order to obtain a measurement for the
radial expansion of the remnant.  The inner ring is inclined by $43\degr$ to
the line of sight \citep{crotts}.  The bulk of the X-ray emission originates
from the ``disk'' containing the inner ring, rather than from a spherical
volume \citep{zhekov05}.  The X-ray morphology of SNR 1987A appears to be an
elliptical ring with 3-4 enhanced lobes (Figure \ref{fig:87a_images}).  The
real number of X-ray spots is likely larger, but the Chandra spatial
resolution prohibits the distinguishing of finer image details.  To
remove the apparent ellipticity prior to fitting the observed images with
our models, we deproject the images by the $43\degr$ inclination.  After
the deprojection, the images are approximately circular (e.g., Figure
\ref{fig:model_fits}a and \ref{fig:model_fits}b).  We fit three simple
empirical models to the image data in an attempt to measure the radial
expansion velocity, with the latter two models removing the biases created by
the brightening hot spots.
  
\subsection{Torus Only Model \label{sec:ring}}
The torus model is the simplest of the three models.  It assumes that the
images can be approximated by a circularly symmetric smooth ring with a peak
radius of $r_0$ and a Gaussian width of $\sigma_{r_0}$, with the following
form:
\begin{equation}\label{eq:r}
  r=\left[(x-x_0)^2+(y-y_0)^2\right]^{1/2}
\end{equation}
\begin{equation}\label{eq:model_torus}
  M_{torus}=N_0+N_r\ exp\left[\frac{-(r-r_0)^2}{2\sigma_{r_0}^2}\right]
\end{equation}
where $x$ and $y$ are the coordinates of each pixel in the model image, $x_0$
and $y_0$ are the offsets of the ring center from the image center, $N_0$ is
the background image level, $N_r$ is the normalization, and $M_{torus}$ is
the resulting model image.  An example of a fit to this model is shown in
Figure \ref{fig:model_fits}c.

\subsection{Bilateral  Lobes + Torus Model \label{sec:lobes}}
This model consists of the torus as described in \S\ref{sec:ring} plus 2
bilateral lobes.  The addition of bilateral lobes approximates the
morphology of the radio remnant \citep{gaensler,manch02,manch05,ng08}.  The
morphology of the X-ray remnant in the earliest epochs resembled that of the
radio remnant with emission approximated by opposite bilateral regions of
increased brightness. As the X-ray spots have brightened, additional lobes
are required to adequately model the images. Therefore, this model works
best in the early epochs and poorly in late ones.  We required the lobes to
be positioned with one 
on each the East and West side of the remnant and allow them to vary in
intensity ($N_{L_i}$), position angle ($\theta_{i}$), and azimuthal
width ($\sigma_{t_i}$). The torus is also allowed to vary in $x_0$,
$y_0$, $r_0$ and $\sigma_{r0}$.  The lobes are then multiplied by the torus
which puts them at a radius of $r_0$ and a radial width of $\sigma_{r0}$,
taking the form:
\begin{equation}\label{eq:l2}
  L_i=N_{L_i}\ exp\left[\frac{-(\theta-\theta_i)^2}{2\sigma_{t_i}}\right]
\end{equation}
where $i$ represents each individual lobe.  We combine the torus and the lobes as:
\begin{equation}\label{eq:model_2lobes}
  M_{2lobes}=M_{torus}(1+L_1+L_2)
\end{equation}
An example of the fit to this model is shown in Figure
\ref{fig:model_fits}d.


\subsection{Quadrilateral Lobes + Torus Model \label{sec:4lobes}}
\begin{figure}
    \begin{center}
      \includegraphics[scale=0.5,angle=90]{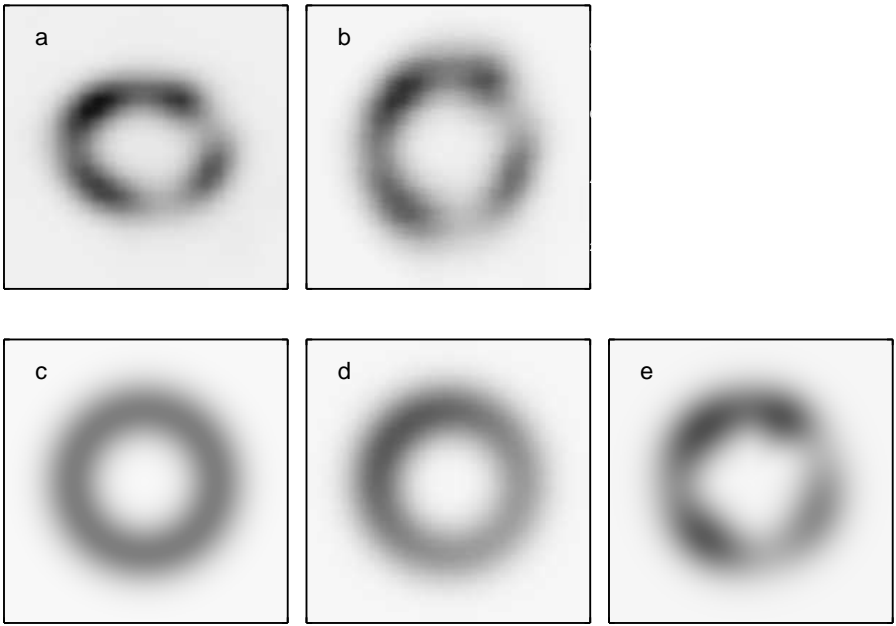}
      \figcaption{Images and fits to models for observation 11.  {\it a}: original
        deconvolved sub-pixel image, {\it b}: deprojected image, {\it c}:
        fit to torus only model, 
        {\it d}: fit to the torus plus two-lobes model, {\it e}: fit to torus plus
        four-lobes model. \label{fig:model_fits}}
    \end{center}
  \end{figure}

A comparison of Figure \ref{fig:model_fits}b and \ref{fig:model_fits}d
(bilateral model) clearly shows that the bilateral model fails to reproduce
the 4 major hot spots seen in the X-ray image.  Because there appears to be
one such hot spot in each quadrant of the image, we next attempt to model
the data using 4 tangentially extended lobes superposed on the torus
described in \ref{sec:ring} with one lobe in each image quadrant (NW, NE,
SE, SW).  However they are
allowed to vary in intensity ($N_{t_i}$), position angle
($\theta_i$) within the quadrant, and azimuthal width
($\sigma_{t_i}$), where $i$ represents the properties of each individual
lobes. The torus is also allowed to vary in $x_0$, 
$y_0$, $r_0$ and $\sigma_{r0}$. The lobes are then multiplied by the torus
which puts them at a radius of $r_0$ and a radial width of $\sigma_{r0}$,
taking the forms of Equation \ref{eq:l2}.
We combine the torus and the lobes as:
\begin{equation}\label{eq:model_lobes}
  M_{4lobes}=M_{torus}(1+L_1+L_2+L_3+L_4)
\end{equation}
An example of the fit to this model is shown in Figure
\ref{fig:model_fits}f.
  
\section{Results \label{sec:results}}

We fit the deprojected images from each epoch to each of the three models
using a Levenberg-Marquardt two-dimensional least-squares IDL fit
procedure.  
Our models are able to sufficiently represent the data in all cases, with
improved reduced $\chi^2$ as model complexity increases.  Example image fits
to all models for sample observation 11 are presented in Figure
\ref{fig:model_fits}.

\begin{figure}
    \begin{center}
      \includegraphics[scale=0.75,angle=90]{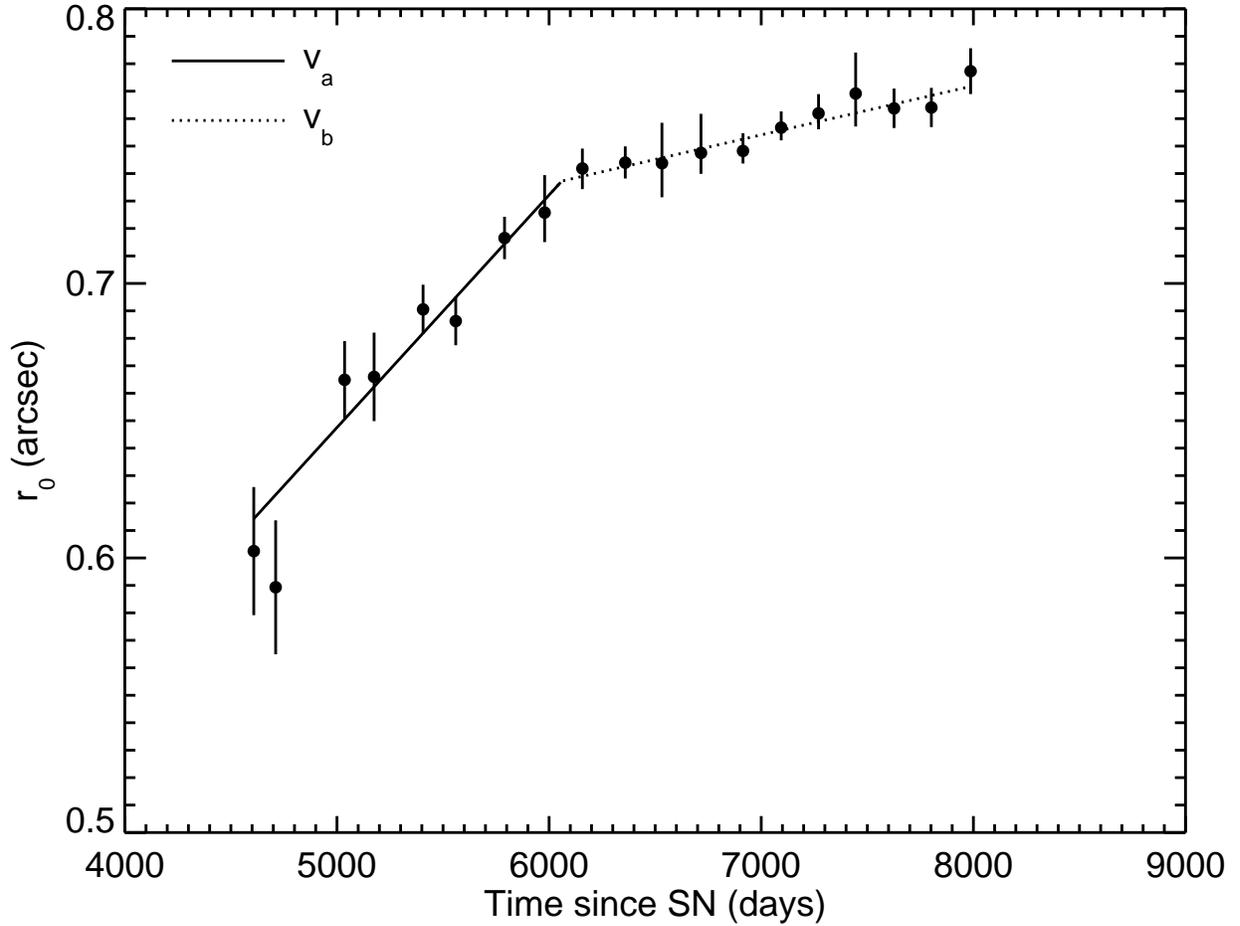}
      \figcaption{Radial expansion measure from the fits to the
        quadrilateral lobes plus torus model for the broken linear fit with
        an early fast shock velocity ($v_a$) transitioning to the later
        slower shock velocity ($v_b$).  
        The radii for the 3 image models are listed in Table
        \ref{tab:rad_detail}, and the resulting radial expansion fits are
        given in Table \ref{tab:results}. 
        \label{fig:tada}}
    \end{center}
  \end{figure}

The main goal of this study is to measure the expansion rate of the remnant
and follow its evolution.  We use the parameter $r_0$ measured in each model
fit to characterize this behavior.  We convert $r_0$ into physical velocity
units using the distance to SNR 1987A ($51.4\pm 1.2$ kpc,
\citealt{panagia03}).  The expansion measures do not increase linearly
throughout, but rather are better fit by a linear fit that extends until
$\sim 6000$ days after the SN, and then turns over to a shallower slope
(Table \ref{tab:rad_detail}, Figure 
\ref{fig:tada}).  Table \ref{tab:results} lists the resulting radial
expansion fit parameters for each set of the image model fits.  All three
models provide consistent fits to the radial expansion measurements with the
tightest constrains on $r_0$ from the model that best represents the images
(quadrilateral lobes plus torus).  The results of these fits are shown Figure
\ref{fig:tada}.

Our estimates of the radial expansion rates before day $6000$ are $\sim65\%$
larger than previous estimates \citep{park04}.  This is most likely due to
the two different methods used to measure the radius of the SNR.  Our radius
estimates use more sophisticated image model fits than the simple Gaussian
radial profile of the X-ray ring used by the previous work, and thus should
be more realistic and reliable.  We also considered the $43\degr$
inclination effect of the inner ring causing it to appear elliptical,
whereas our previous work assumed a 
simple circular geometry.  For comparison, if we re-incline the measured
radial expansion rate, the corresponding expansion velocity is $\sim
5500^{+1600}_{-1100} \kms$ prior to day $\sim 6000$ and $\sim 1200
^{+460}_{-440} \kms$ afterwards.  This is in plausible agreement with the previous
measurements.  Hydrodynamic models indicate a forward shock velocity of
$\sim3500-4500\kms$ until day $\sim6000$, and then a dramatic deceleration
down to $\sim1000-1500\kms$ afterwards \citep{dwarkadas07}.  This overall
change of the shock velocity is also in good agreement with the results from
our expansion rate estimates.
  
The most recent radio radial expansion measurements by \cite{ng08} show a
constant remnant expansion rate of $4000\pm400\kms$ over a range of
1992-2008.  However, the 
radio remnant radius is systematically larger than the X-ray remnant radius
by $\sim 10-20\%$. Interestingly, after day $\sim 5400$, the radio
remnant radius is larger than the radius of the optical inner ring ($0\farcs
83$, \citealt{sugerman02}).  This discrepancy in the SNR size between the X-ray and
radio data is largely caused by the different image modeling techniques as
discussed by \cite{ng08}.  It may also indicate that the X-ray and
radio emission are produced in different physical regions.  The radio
emission is likely due to synchrotron processes in between the forward and
reverse shocks \citep{manch05}, while the X-rays are dominated by thermal
emission from the transmitted and reflected shocks produced by the
shock-inner ring interaction \citep{zhekov09}.  The 3-dimensional effects
and emission regions also likely differ between these components.
\cite{ng08} predict that they will soon see a deceleration in the
radio remnant expansion similar to that we have observed from the X-ray
remnant due to the forward shock interacting with the optical inner ring.

  
  
\section{Exploring Methods with Simulated Images \label{sec:sims}}

In order to test the reliability of our image modeling and radial expansion
measures, we created a series of simulated images using the {\it Model of AXAF
Response to X-rays} (MARX) tool.
The simulated images were meant to approximate the real images obtained at
days $\sim 4600$, 5200, 6200, and 7200 days after the SN explosion, assuming
the estimated radius for each epoch.  They were generated with different
observed inclinations and properties, specifically 0 degrees inclination
(face-on), and 45 degrees inclination both with and without 4 spots.  We
applied the {\it Chandra} PSF to the simulated images to mimic observing
effects, and ran them through the same image deconvolution process as done
with the real data.

We processed these simulated images through the same algorithms used to fit
the observed images for all of our image models.  The results showed a
systematic offset in the best-fit radius, which is $\sim 10\%$ smaller than
the assumed input values, yet gave an accurate measure of the radial
expansion rate within measurement errors.  We attribute this systematic
offset to differences in the treatment of the torus width ($\sigma_{r_0}$).
The simulated images 
were constructed assuming that the torus and any changes to the width over
time are not spatially resolved.  Applying this same assumption to the
simulation fits, we froze the torus width to approximately the value of the
effective PSF for the deconvolved images ($\sim 0\farcs 2-0\farcs 3$), and
redid the fits.  This effectively eliminated the systematic offset and
reproduced the input radii accurately.

In the application to the real data, when leaving the torus width as a free
parameter, the model fits may underestimate the radius of the real remnant
images by as much as $10\%$.  However, the free torus width fits appear to
reveal a widening torus with values at a similar size to the PSF.
Any real physical changes to the torus width would be ignored if the
parameter is frozen to a set value.  The width is not independent of the
radii measurement that we are inherently trying to measure.  
The fits to the real data are statistically improved if we
assume that the torus has some real width that may change over time and
the width parameter is left free to vary.  The fitted torus width (Gaussian
half-width) ranges from $\sim 0\farcs 12-0\farcs 21$ which is significantly
smaller than the ring radius and comparable to the effective PSF of the
deconvolved {\it Chandra} images.

We also assume a 2-dimensional structure in all of our model fitting even
with the deprojecting correction.  In reality, X-ray emission most likely
originates from a 3-dimensional structure with the shock component being
perpendicular to the disk containing the inner ring, making contributions to
the torus width unknown.  The real 3-dimensional geometry of the unresolved
inner ring is not yet well understood.  The potential $\sim
10\%$ radii offset is systematic and may consequently have the same
effect on the expansion velocities.  However, the measurement errors are
larger than this systematic offset.

The first two images are most strongly affected by freezing the
torus width because they suffer from low count statistics (lower by
an order of magnitude than those at other epochs), and
contain the lowest contributions of the X-ray emission from the shock-inner
ring interaction.  Therefore, this 3-dimensional effect could have been
relatively large in these early images.
To test this effect, we fit the radial expansion rate using the frozen torus
width with and without the first two images, with the latter analysis
leading to a more accurate reproduction of the results from the free torus
width fits.  When including the first two observations with a frozen
torus radius ($r_0\sim 0.15\arcsec$), the initial expansion velocity is $\sim
12000 \kms$, flattening to a velocity of $\sim 1400 \kms$ at $\sim 5800$
days.  The expansion measurement excluding the first two observations yields
an initial expansion velocity of $\sim 6000 \kms$ flattening to a velocity
of $\sim 1700 \kms$ at $\sim 6200$ days.  Therefore, it is these early
images which are influencing the fits and are least conducive to a
frozen torus width.  Even with a free torus width, the first two images are
the noisiest and provide the poorest fits.  The first two images have some
effect on the accuracy of the expansion rate, but do not change the overall
conclusion that we observe a slowing of the expansion rate starting at $\sim
6000$ days. 

\section{Conclusions \label{sec:conclusions}}


Multi-wavelength studies at frequent epochs are carefully capturing the
evolution and progression of SNR 1987A.  As the evolution continues, it
reveals new regions of emission and probes a larger volume of the
circumstellar environment.  We believe that the predicted events
\citep{cheva89,cheva95,luo91,bork} of the SN blast wave engulfing the inner
circumstellar ring are occurring.  The earlier optical and soft X-ray
emission spots were a result of higher density gas protruding inward from
the main ring. In recent years, we have seen the entire ring begin to light
up.  The current optical and X-ray emission is a composite of the shocks
heating the higher and lower density regions of the clumpy ring structure.
  
Our measurement of the deceleration of the radial expansion of SNR 1987A
from $\sim 8000 \kms$ to $\sim 1600\kms$ at $\sim 6000$ days along with the 
upturn of the X-ray luminosity, fractional contribution from the decelerated
shock to the soft X-ray flux, and the low shock velocities from line
broadening are all pointing to indications of the SN blast wave encountering
the main body of the circumstellar ring around day 6200.  The evidence of
the blast wave encountering the inner ring are already apparent in the X-ray
observations, and we are yet to see the strong indications from the radio or
optical observations.  The details of the different emission regions and
their interactions with the different shock components will become clearer
as continued optical, X-ray, and radio observations probe the remnant.

Continued monitoring of SNR 1987A with {\it Chandra} will reveal more
details in the structure of the circumstellar environment around the
remnant.  We expect to see additional brightening as the blast wave shock
front heats the material in the ring and as the reverse shock region
continues to grow in volume.  As the remnant expands and brightens, we hope
to further resolve structures and to understand the inhomogeneous nature of
the circumstellar environment.  We expect the expansion to continue to slow
as it sweeps up more material and transitions from the free expansion phase
to the adiabatic expansion phase once it has swept up a mass of material
comparable to the SN ejecta.  Uncertainties in this ejecta mass make
predictions for this time frame difficult.  However, SNR 1987A has an
unusually high density environment as evident by the effects on the original
SN explosion.  Therefore this transition may occur earlier than the
typical $\sim200$ year timescale.
  
We may also expect to see additional deceleration of the radial expansion
while the blast wave sweeps up more of the circumstellar material.  Only the
material that was photo-ionized by the UV flash from the SN is visible in
the optical inner ring, and
there is evidence suggesting that there may be more mass that was not
ionized \citep{crotts89,ensman92}.  The inner ring has an ionized mass of
$\sim 0.04\rm{M_{\odot}}$, which 
is probably only the inner surface of a larger gas complex.  As the hot
spots brighten and produce ionizing radiation, eventually the undetected
outer material will become visible as an emission nebula \citep{mccray03}.
The UV and optical light echoes seen several months after the SN event
hinted at the structure of the dust in the outer circumstellar
environment. It will be interesting to continue to use the radial expansion
measurement to probe the density structure of the ring.
  

\acknowledgements This work is supported by Chandra SAO grants
GO7-8069X and GO8-9076X.  S.P. was also supported in part by SAO grant
SV4-74018.

\clearpage
\bibliographystyle{apj} 
\bibliography{87a_radexp_v6}

  
\clearpage
\renewcommand{\thefootnote}{\fnsymbol{footnote}}
\begin{deluxetable}{clcccl}
  \tablecolumns{6} \tablewidth{0pc} \tablecaption{Chandra Observations of
    SNR 1987A\label{tab:obs_detail}} \tablehead{
    \colhead{Observation \#} & \colhead{Obs Date} &
    \colhead{Age\tablenotemark{*}} & \colhead{Instrument\tablenotemark{\diamond}} & \colhead{Exposure} & \colhead{Source} \\
    \colhead{} & \colhead{} & \colhead{} & \colhead{} & \colhead{(ks)} &
    \colhead{Counts\tablenotemark{\ddagger}} } \startdata
  1 & 1999 October 6 & 4608 & ACIS-S + HETG\tablenotemark{\dag} (3.1 s)& 116.1 & 690\\
  2 & 2000 January 17 & 4711 & ACIS-S3  (3.2 s) & 8.6 & 607\\
  3 & 2000 December 7 & 5036 &  ACIS-S3 (3.2 s) & 98.8 & 9030\\
  4 & 2001 April  25 & 5175 & ACIS-S3  (3.2 s) & 17.8 & 1800 \\
  5 & 2001 December 12 & 5406 &  ACIS-S3  (3.1 s) & 49.4 & 6226\\
  6 & 2002 May 5 & 5560 &  ACIS-S3 (3.1 s) & 44.3 & 6427\\
  7 & 2002 December 31 & 5790 &  ACIS-S3 (3.1 s) & 49.0 & 9277\\
  8 & 2003 July 8 & 5979 &  ACIS-S3 (3.1 s) & 45.3 & 9668\\
  9 & 2004 January 2 & 6157 &  ACIS-S3 (3.1 s) & 46.5 & 11856\\
  10 & 2004 July 22 & 6359 &  ACIS-S3 (1.5 s) & 48.8 & 17979\\
  11 & 2005 January 12 & 6532 & ACIS-S3 (0.4 s) & 48.3 & 24939 \\
  12 & 2005 July 14 & 6716 & ACIS-S3 (0.4 s) & 44.1 & 27048\\
  13 & 2006 January 28 & 6914 & ACIS-S3 (0.4 s) & 42.3 & 30940 \\
  14 & 2006 July 28 & 7094 & ACIS-S3 (0.4 s) & 36.4 & 30870 \\
  15 & 2007 January 19 & 7270 & ACIS-S3 (0.4 s) & 33.5 & 32798 \\
  16 & 2007 July 13 & 7445 & ACIS-S3 (0.4 s) & 25.7 & 27945 \\
  17 & 2008 January 10 & 7626 & ACIS-S3 (0.2 s) & 9.3 & 12008 \\
  18 & 2008 July 4 & 7802 & ACIS-S3 (0.2 s) & 8.6 & 12119 \\
  19 & 2009 January 5 & 7987 & ACIS-S3 (0.2 s) & 6.0 & 9204 \\
  \enddata
  \tablenotetext{*}{Days since SN explosion} 
  \tablenotetext{\dag}{Zeroth order data from grating observations}
  \tablenotetext{\ddagger}{$0.3-8.0$ keV band}
  \tablenotetext{\diamond}{Time in parentheses is the ACIS frame-time.}
\end{deluxetable}
  
\begin{deluxetable}{lllllll}
  \tabletypesize{\small}
  \tablecolumns{7} \tablewidth{0pc} \tablecaption{Model Parameter
    Fits\label{tab:rad_detail}} \tablehead{  
    \colhead{Age\tablenotemark{*}} &
    \multicolumn{2}{c}{Torus Only} & \multicolumn{2}{c}{Torus+2 lobes} &
    \multicolumn{2}{c}{Torus+4 lobes} \\
    \colhead{} & \colhead{$\sigma_{r_0}$ (arcsec)} & \colhead{$r_0$
      (arcsec)} & \colhead{$\sigma_{r_0}$ (arcsec)} &  \colhead{$r_0$ (arcsec)}
    & \colhead{$\sigma_{r_0}$ (arcsec)} & \colhead{$r_0$ (arcsec)}}
  \startdata
4608 & $0.12 \pm 0.01$ & $0.60 \pm 0.02$  & $0.12 \pm 0.01$ & $0.60 \pm 0.02$  & $0.12 \pm 0.01$ & $0.60 \pm 0.02$  \\
4711 & $0.14 \pm 0.01$ & $0.59 \pm 0.02$  & $0.14 \pm 0.01$ & $0.59 \pm 0.02$  & $0.14 \pm 0.01$ & $0.59 \pm 0.02$  \\
5036 & $0.19 \pm 0.01$ & $0.65 \pm 0.01$  & $0.18 \pm 0.01$ & $0.66 \pm 0.01$  & $0.18 \pm 0.01$ & $0.66 \pm 0.01$  \\
5175 & $0.17 \pm 0.01$ & $0.66 \pm 0.02$  & $0.16 \pm 0.01$ & $0.64 \pm 0.02$  & $0.17 \pm 0.01$ & $0.67 \pm 0.02$  \\
5406 & $0.19 \pm 0.01$ & $0.68 \pm 0.01$  & $0.19 \pm 0.01$ & $0.68 \pm 0.01$  & $0.19 \pm 0.01$ & $0.69 \pm 0.01$  \\
5560 & $0.19 \pm 0.01$ & $0.68 \pm 0.01$  & $0.19 \pm 0.01$ & $0.67 \pm 0.01$  & $0.19 \pm 0.01$ & $0.69 \pm 0.01$  \\
5790 & $0.22 \pm 0.01$ & $0.69 \pm 0.01$  & $0.22 \pm 0.01$ & $0.71 \pm 0.01$  & $0.22 \pm 0.01$ & $0.72 \pm 0.01$  \\
5979 & $0.20 \pm 0.01$ & $0.73 \pm 0.01$  & $0.21 \pm 0.01$ & $0.72 \pm 0.01$  & $0.21 \pm 0.01$ & $0.73 \pm 0.01$  \\
6157 & $0.22 \pm 0.01$ & $0.73 \pm 0.01$  & $0.22 \pm 0.01$ & $0.73 \pm 0.01$  & $0.22 \pm 0.01$ & $0.74 \pm 0.01$  \\
6359 & $0.22 \pm 0.01$ & $0.74 \pm 0.01$  & $0.22 \pm 0.01$ & $0.74 \pm 0.01$  & $0.22 \pm 0.01$ & $0.74 \pm 0.01$  \\
6532 & $0.22 \pm 0.01$ & $0.74 \pm 0.01$  & $0.21 \pm 0.01$ & $0.74 \pm 0.01$  & $0.21 \pm 0.01$ & $0.74 \pm 0.01$  \\
6716 & $0.21 \pm 0.01$ & $0.75 \pm 0.01$  & $0.21 \pm 0.01$ & $0.74 \pm 0.01$  & $0.21 \pm 0.01$ & $0.75 \pm 0.01$  \\
6914 & $0.22 \pm 0.01$ & $0.74 \pm 0.01$  & $0.22 \pm 0.01$ & $0.74 \pm 0.01$  & $0.22 \pm 0.01$ & $0.75 \pm 0.01$  \\
7094 & $0.21 \pm 0.01$ & $0.76 \pm 0.01$  & $0.21 \pm 0.01$ & $0.75 \pm 0.01$  & $0.21 \pm 0.01$ & $0.76 \pm 0.01$  \\
7270 & $0.21 \pm 0.01$ & $0.76 \pm 0.01$  & $0.21 \pm 0.01$ & $0.76 \pm 0.01$  & $0.21 \pm 0.01$ & $0.76 \pm 0.01$  \\
7445 & $0.20 \pm 0.01$ & $0.77 \pm 0.01$  & $0.21 \pm 0.01$ & $0.77 \pm 0.01$  & $0.21 \pm 0.01$ & $0.77 \pm 0.01$  \\
7626 & $0.21 \pm 0.01$ & $0.76 \pm 0.01$  & $0.21 \pm 0.01$ & $0.76 \pm 0.01$  & $0.21 \pm 0.01$ & $0.76 \pm 0.01$  \\
7802 & $0.20 \pm 0.01$ & $0.77 \pm 0.01$  & $0.20 \pm 0.01$ & $0.76 \pm 0.01$  & $0.20 \pm 0.01$ & $0.76 \pm 0.01$  \\
7987 & $0.20 \pm 0.01$ & $0.78 \pm 0.01$  & $0.20 \pm 0.01$ & $0.78 \pm 0.01$  & $0.20 \pm 0.01$ & $0.78 \pm 0.01$  \\

  \enddata
   \tablenotetext{*}{Days since SN explosion} 
   \tablecomments{Errors are 1$\sigma$ confidence intervals.}
\end{deluxetable}

\begin{deluxetable}{ccccc}
  \tablecolumns{6} \tablewidth{0pc} \tablecaption{Radial Expansion
    Velocity Fits \label{tab:results}} \tablehead{ \colhead{} &
    \colhead{$v_a$} & \colhead{turn over time} & \colhead{$v_b$} &
    \colhead{$\chi^2_{v_{ab}}$/dof} \\
    \colhead{} & \colhead{($\rm{km}\,\rm{s}^{-1}$)} & \colhead{(days)} &
    \colhead{($\rm{km}\,\rm{s}^{-1}$)} & \colhead{}} 
  \startdata Torus only & $6908_{-1127}^{+1283}$ & $6184_{-220}^{+256}$ & $1961_{-635}^{+602}$ & $0.70$ \\
  Torus + 2 Lobes & $7047_{-1204}^{+1460}$ & $6135_{-284}^{+208}$ & $2037_{-578}^{+532}$ & $0.53$ \\
  Torus + 4 Lobes & $7539_{-1524}^{+2139}$ & $6060_{-237}^{+208}$ & $1592_{-566}^{+589}$ & $0.54$ \\
  
  \enddata
  \tablecomments{$v_a$ and $v_b$ are calculated using the slopes of the
    broken linear fit with 15 DOF.  All
    errors are 90$\%$ confidence.}
\end{deluxetable}
    
  \clearpage
  
\end{document}